\documentclass[aps,prl,twocolumn,showpacs]{revtex4}
\usepackage{epsfig}
\usepackage{graphicx}
\usepackage{bm}
\usepackage{amssymb}
\usepackage{amsmath}
\usepackage{amsthm}
\begin{document}
\title{Fluctuations in fluid invasion into disordered media}
\author{Martin Rost$^1$, Lasse Laurson$^2$, Martin Dub\'e$^3$, Mikko
  Alava$^2$} 
\affiliation{$^1$ Theoretische Biologie, IZMB, Universit\"at Bonn,
  53115 Bonn, Germany\\
  $^2$ Physics Laboratory, Helsinki University of Technology,
  P.L.~1100, 02105 HUT, Espoo, Finland\\
  $^3$ CIPP, Universit\'e de Qu\'ebec \`a Trois-Rivi\`eres, C.P.~500,
  Trois-Rivi\`eres, Qu\'ebec, G9A 5H7 Canada}
\begin{abstract}
Interfaces moving in a disordered medium exhibit stochastic velocity
fluctuations obeying universal scaling relations related to the
presence or absence of conservation laws. For fluid invasion of porous 
media, we show that the fluctuations of the velocity are governed by a
geometry-dependent length scale arising from fluid conservation. This
result is compared to the statistics resulting from a 
non-equilibrium (depinning) transition between a moving interface and
a stationary, pinned one.

\end{abstract}
\pacs{05.40.-a, 47.56.+r, 47.61.Jd, 68.35.Ct}
\maketitle

The dynamics of interfaces, domain-walls and fronts in disordered
systems has attracted tremendous attention not only because of
practical applications but also for the general interest in the
emerging equations of motion and scaling laws. The presence of
randomness is felt through both frozen (or ``quenched'') noise and
``thermal'' fluctuations. Experimentally tested scenarios include fire
fronts in combustible media \cite{mau97}, cracks propagating in solids
\cite{mal06,sal06,kun04}, domain walls in magnets
\cite{mei92,heb02,har04,dur06,par05}, and interfaces in multi-phase
flows (``imbibition'' or ``capillary rise'' phenomena)
\cite{sor02,ger02,sor05,sor02b,dou98}, to name a few.

In these examples the fluctuations of the fronts can generally be
described via self-affine fractals, with dynamics exhibiting critical
correlations. For time-independent randomness, such as the pore
structure in fluid invasion, it is possible to concentrate on the
zero-temperature phase-transition which separates moving interfaces
from pinned ones. The situation is then represented in terms of an
order parameter, the velocity of the interface, and a control
parameter, the driving force. The interface is described by a height
$h({\bf x},t)$, at position ${\bf x} \in \mathbb{R}^d$ and time $t$,
with its dynamics given by a Langevin equation $\partial_t h \! = \!
{\cal L} (\{ h \}, \alpha(\{h\},{\bf x},t))$. The kernel ${\cal L}$
contains the deterministic contribution of the interface and the
random contribution of the noise configuration $\alpha(\{h\},{\bf
  x},t)$. Both may depend non--locally on the entire interface
configuration, denoted by $\{h\}$.

The scaling exponents of $h({\bf x},t)$ are often studied but are by
no means the only interesting feature of this problem. It is also
possible to study the amplitude of the various correlation functions,
together with the persistence properties of the interface
\cite{burning}, or the scaling of probability distributions of
quantities such as deviations of the interface from its average
position or velocity.

Here, we study the velocity fluctuations of driven fluid interfaces in
disordered porous media in an imbibition situation, where a viscous
fluid displaces air or a  less viscous fluid. It is an important
problem in the general field of fluid flow in porous media, with many
engineering and technological applications. Fluid invasion and
multi-phase flows are often analyzed  via lattice models of pore
networks (e.g.\ for oil industry applications), or via ``upscaling
techniques'', which summarize the physics on a coarse-grained (volume
element) scale. One often studied aspect is the averaging of the
scale-dependent permeability $\kappa({\bf r})$, from the smallest pore
scale up to geological scales \cite{bra05,noe05,gas05,ahm96}. It is
not trivial to arrive to a coarse-grained description of these
systems, since the physics is governed by the (conserved) fluid flow
in the porous bulk coupled to microscopic interfaces, menisci, between
the two fluids. 

We report on the behavior of the spatially averaged interface velocity
$v(t)$, directly related to the liquid intake under an externally
controlled pressure. We first show that there exists a scaling
relation between  the velocity fluctuations $\Delta v$ and the average
velocity $\bar v = \langle v(t) \rangle$. This scaling is also
reflected in the power spectrum of $v(t)$ as we discuss in detail.
The scaling arguments are backed with numerical simulations of a
phase-field model for imbibition \cite{dub99}. We also study the
interface velocity fluctuations for non-conserved, ``local'' dynamics,
where we find a {\em different} scaling relation, again backed by
simulations of the appropriate model \cite{les93}. Both the average
flux as well as its fluctuations are experimentally well accessible so
our theoretical approach allows for an easy assessment of the type of
dynamics. The detailed experiments of the Barcelona group and others
point into that direction \cite{sor02,ger02,sor02b,sor05,dou98}, and
the main result extends to other systems where the front invasion
depends on a conserved quantity.

The interface advances because fluid is transported from the reservoir
through the medium. By Darcy's law, fluid flow is proportional to a
pressure gradient,
\begin{equation}
{\bf j} = \frac{\kappa}{\eta} \; \nabla P.
\label{eq_darcy}
\end{equation}
Here $\eta$ is the viscosity of the liquid and $\kappa$ the
permeability of the medium. Deviations from Eq.~(\ref{eq_darcy}) are
known to arise for microscopic reasons, such as, e.g., inertial effects 
when a meniscus enters a pore, clogging by advected particles as in
the case of a dye, and the partial wetting of
capillaries. Nevertheless, the linear form already leads to rich
phenomena \cite{ala04}.

For an incompressible fluid, 
a Laplace--equation $\nabla^2 P = 0$ governs the pressure field in the
bulk. The appropriate boundary conditions are  (i) $P \equiv
P_{\rm R}$ at the contact to a reservoir of liquid with constant
pressure, and (ii) $P_{\rm I} \! = \! \gamma^* {\cal K} \! + \! P_c \!
+ \! P_0$ at the interface. Condition (ii) is a superposition of
atmospheric ($P_0$) and capillary ($P_c$) pressure on the pore scale
and the effect of curvature ${\cal K}$ on a {\em coarse-grained} scale
by an effective interface tension  $\gamma^*$. For weak disorder, as
arise in e.g.\ paper, it is
close to the bare value $\gamma$, but may be much
smaller in porous materials with weakly connected channels such as
Vycor glass \cite{hub04}.

For $P_{\rm R} \! = \! P_0$ imbibition is spontaneous with an average
velocity $\bar{v} \! \equiv \! \partial_t \bar h \! = \! (\kappa/\eta)
\; P_c/\bar h$, where $\bar h(t)$ is the average height. This leads to
Washburn's law of capillary rise, $\bar h \! \sim \! t^{1/2}$
\cite{was21}. Forced imbibition arises by increasing the pressure at
the reservoir as the interface advances. A constant mean pressure
gradient, $|\nabla P| \! = \! (P_{\rm R} \! - \! P_{\rm I})/\bar h \!
= \! \eta \bar{v} / \kappa$, keeps the average velocity $\bar v$
constant.

Quenched disorder in the porous structure roughens the interface
around its average position. Capillary disorder $p_c ({\bf r}) \! = \!
P_c \! + \! \delta p_c({\bf r})$ acts only at the interface, and
permeability disorder, $\kappa({\bf r}) \! = \! \kappa \! + \! \delta
\kappa({\bf r})$ controls the flux of liquid from the reservoir to the
interface. If $r_0$ is the typical pore size and $\delta r_0$ its
deviation, then $p_c \! \sim \! \gamma^* / r_0$ and $\kappa \! \sim \!
r_0^2$, with $\delta p_c /\delta \kappa \! \sim \! p_c/\kappa$
\cite{pau03,ala04}.

Effective (capillary) interface tension and the average pressure
gradient smoothen the interface and allow for correlated roughness
only up to a lateral length scale
\begin{equation}
\xi_c \sim \sqrt{\frac{\kappa \gamma^*}{\eta \bar v}} =
\sqrt{\frac{\kappa}{\rm Ca}}
\label{xi}
\end{equation}
related to the capillary number ${\rm Ca} \! = \! \eta \bar
v/\gamma^*$ \cite{dub99,ala04}. Its presence and dependence on $\bar
v$ has also been confirmed experimentally \cite{sor02b,ger02}. The
ratio of disorder strengths imposes another length scale
\begin{equation}
\xi_\kappa \sim \frac{\kappa^2}{\bar v \eta} \; \frac{\delta
p_c}{\delta \kappa}  = \frac{\sqrt{\kappa}}{\rm Ca}
\label{xikappa}
\end{equation}
separating regimes of different roughness. On scales shorter than
$\xi_\kappa$ interface roughness is caused by capillary disorder, on
larger scales by permeability disorder \cite{pau03,ala04}. We consider
the interesting case of a slowly advancing front, where $\xi_\kappa \!
> \! \xi_c$, and capillarity induced fluctuations prevail, and one has
a well defined interface (for stability a ${\rm Ca}$ below
$\sim {\cal O}(1)$ is needed \cite{len90}). Then, permeablity noise
and its possible correlations with capillary disorder can be ignored.

The scaling behavior in imbibition can be demonstrated by model
simulations \cite{dub99}. A scalar phase field $\phi({\bf r},t)$
denotes by $\phi \! = \! 1$ the invaded (wet) and $- \! 1$ the
non-invaded (dry) regions respectively. An energy functional
\begin{equation}
{\cal F}[\phi] = \int d^{d+1}r \left[ \frac{(\nabla \phi)^2}{2} -
  \frac{\phi^2}{2} + \frac{\phi^4}{4} - \alpha({\bf r}) \phi \right]
\end{equation}
couples the phase field to a space dependent quenched randomness
$\alpha({\bf r})$ with average $\bar \alpha$ and standard deviation
$\Delta \alpha$. This term models capillarity disorder, with sign
($\alpha \! > \! 0$) favoring the wet phase $\phi \! = \! 1$. The
model is defined on the half--space, ${\bf r} = ({\bf x},y)$ with $y
\! > \! 0$, with boundary conditions $\phi \! \equiv \! 1$ at $y \! =
\! 0$ (coupled to a wet reservoir at the bottom) and initial condition
$\phi({\bf r},t \! = \! 0) \! \equiv \! - \! 1$ (dry).

\begin{figure}[!h]
\begin{center}
\includegraphics[width = 3.8cm]{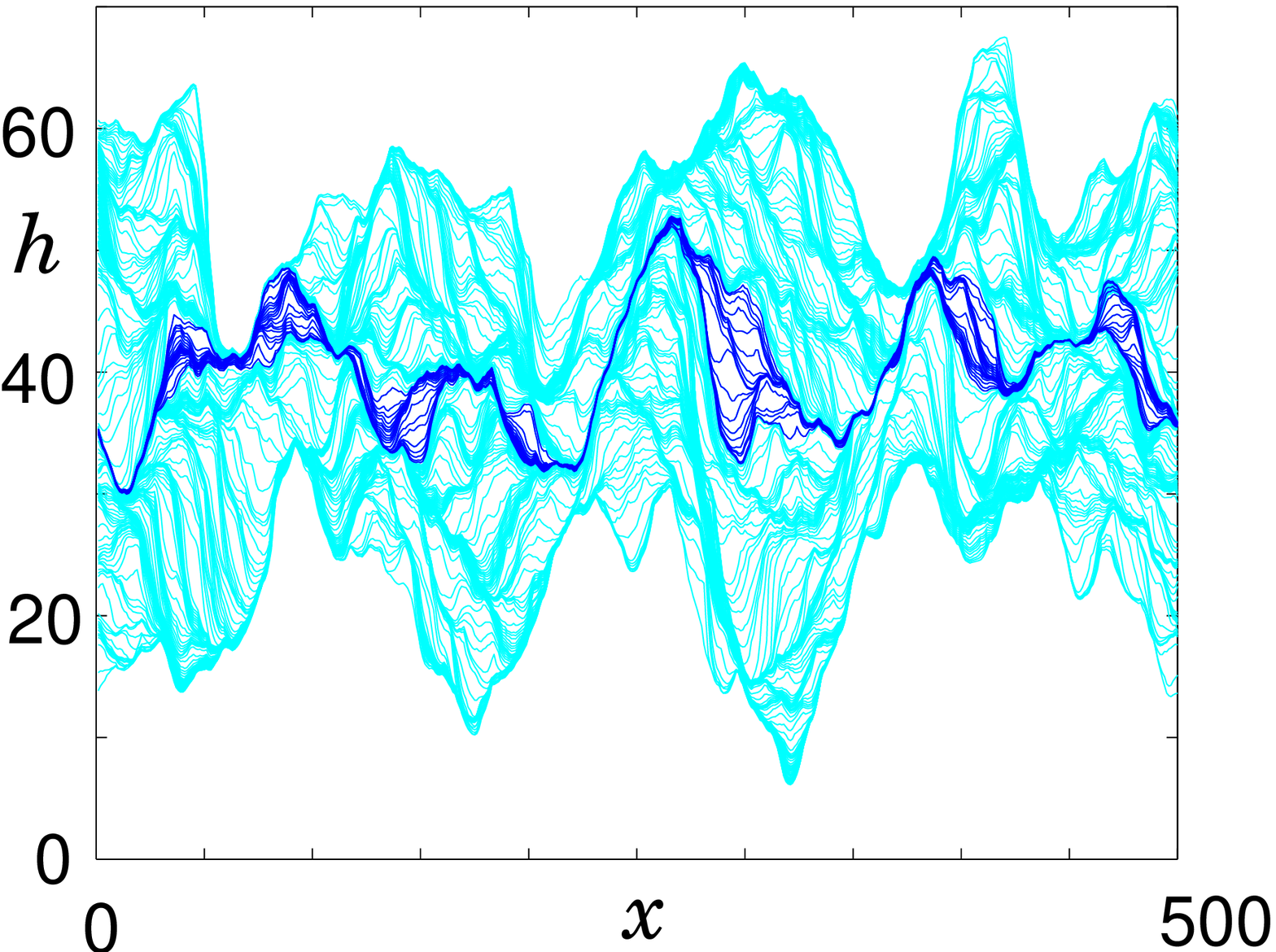}
\includegraphics[width = 4.3cm]{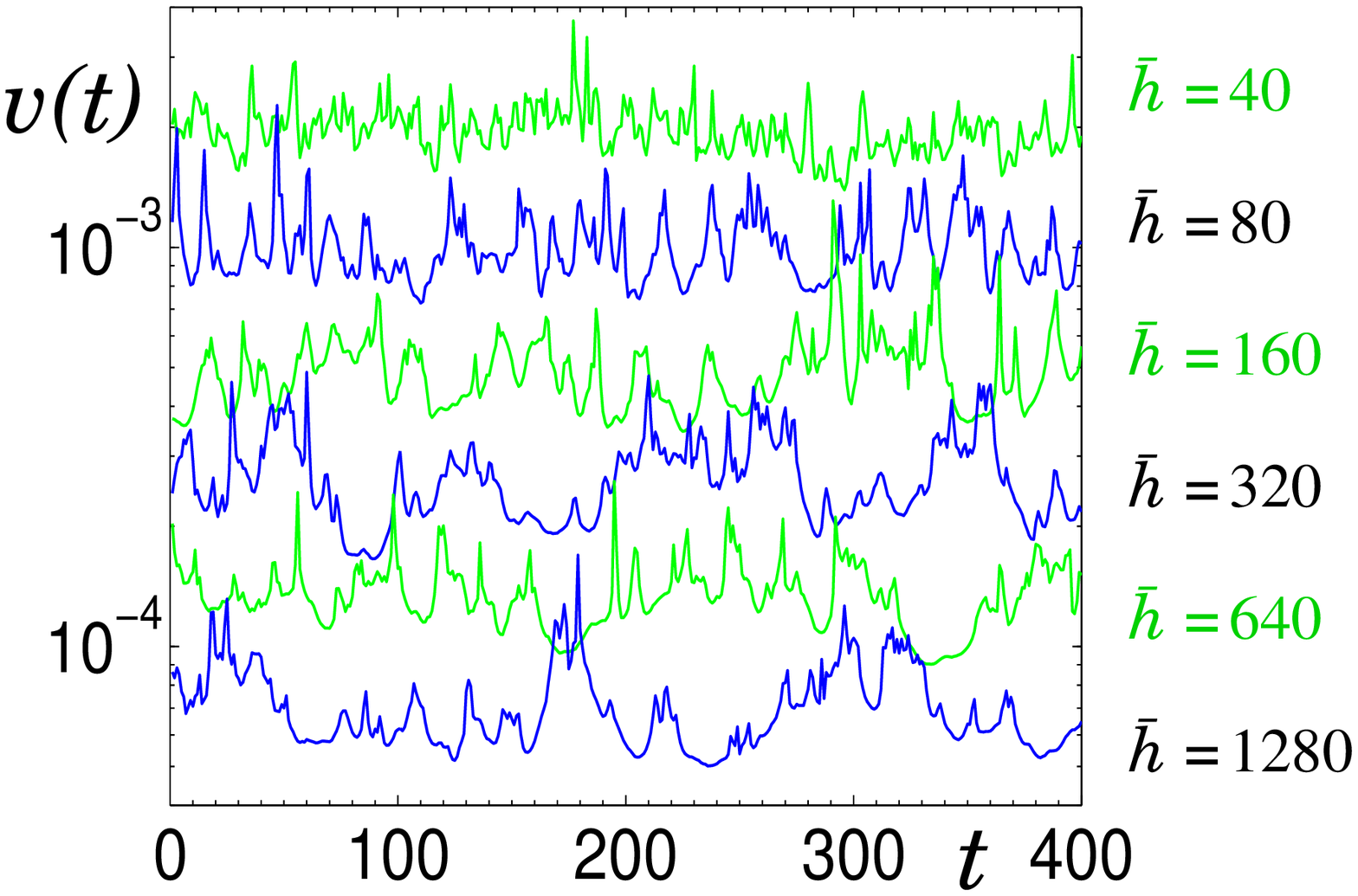}
\end{center}
\caption{{\bf Left}: 
  Interfaces $h(x,t)$ in a system of width $L = 500$ and average
  separation $\bar h = 320$ from the reservoir. The configurations are
  separated by time intervals $\Delta t = 500$ at 240 different
  times. Interface propagation is made visible by adding the constant
  shift velocity $\bar v$. The sequence of 20 black interfaces shows
  pinned and moving regions.
  {\bf Right}:
  time series of $v(t)$, the velocity for $L = \bar h = 40$, $80$,
  $160 \dots 1280$ from top to bottom, logarithmic scale in $v(t)$.}
\label{sample}
\end{figure}

\begin{figure}[!h]
\vspace{-.35cm}
\begin{center}
\includegraphics[width = 7cm]{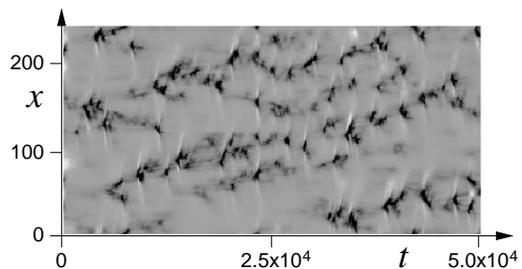}
\end{center}
\vspace{-.8cm}
\caption{Activity pattern in space (vertical) and time
  (horizontal). $L = 256$, $\bar h = 160$, lateral periodic boundary
  conditions. High velocities dark, low velocities light, nonlinear
  grey--scale for better visibility. At any given time activity is
  restricted to $n = L/ \bar h \approx 2$ narrow avalanches of width
  $\xi_c$.}
\label{fig:act}
\end{figure}

The corresponding chemical potential $\mu \! = \! \delta {\cal
  F}/\delta \phi$ (with the role of pressure in the context of
imbibition) drives a current ${\bf j} = - \tilde \kappa \nabla \mu$
with a rescaled and possibly spatially varying mobility $\tilde
\kappa({\bf r})$. By local conservation of $\phi$ this yields the
dynamical equation
\begin{equation}
\partial_t \phi = - \nabla \cdot \tilde \kappa({\bf r}) \; \nabla
  \left[ \nabla^2 \phi + \phi - \phi^3 + \alpha({\bf r}) \right].
\label{eqofmot}
\end{equation}
With this model, a Young--Laplace relation between the chemical
potential at the interface and the curvature, $\mu_{\rm int} \! = \!
{\cal K} \! - \! \alpha$ arises naturally. 

The numerical simulations are performed, in a way that mimics
forced--flow imbibition, by continuously shifting the fields
$\phi({\bf r},t)$, $\tilde \kappa({\bf r})$ and $\alpha({\bf r})$
downward at velocity $\bar v$, as in the experiments of
Ref.~\cite{hor93}. The stationary interface then has average height
$\bar h \! = \! \bar \alpha / (2 \bar v)$. For the ${\rm Ca}$ at hand
the permeability disorder has no influence (see \cite{ala04}), nor
have correlations between $\tilde \kappa({\bf r})$ and $\alpha({\bf
  r})$. Only simulations with capillary disorder are thus discussed
here. Fig.~\ref{sample} shows examples of model interfaces in $d \! +
\! 1 \! = \! 2$ dimensions, together with timeseries of average
velocity or liquid intake, $v(t)$ exhibiting typical
fluctuations. Fig.~\ref{fig:act} shows a spatiotemporal activity
pattern of the local interface velocities $\partial_t h(x,t)$.

The {\em dynamics} of the roughening interface can be understood as
a sequence of avalanches. At any given time only a small localized
portion of the interface is moving fast. There the Laplacian pressure
field drags liquid from a surrounding region of {\em lateral} size
${\bar h}^d$, region which therefore remains pinned. Further away the
interface is not affected and moves independently. Within this region,
the moving part has {\em maximal} capillary forces $p_c$. However, it
eventually encounters disorder with lower $p_c$ and gets pinned. At
this point, another region elsewhere starts to move. Propagation by
avalanches in imbibition and their size distribution with one fixed
scale has been found and discussed in experiments \cite{dou98}. It is
visible in Figs.~\ref{sample} and \ref{fig:act}. The size of these
moving avalanches is given by the cutoff lengthscale $\xi \! = \!
\xi_c$ or $\xi_\kappa$. The regime governed by capillary disorder has
global roughness exponents $\chi \! \approx \! 1.25$ and a different
local exponent $\chi_{\rm loc} \! = \! 1$  in $d \! + \! 1 \! = \! 2$
dimensions \cite{dub99,ala04,lop97}. For higher ${\rm Ca}$ than those
considered here, permeability disorder starts to play a role and the
exponents are $\chi \! = \! \chi_{\rm loc} \! = \! 1$ in $d \! + \! 1
\! = \! 2$ dimensions \cite{pau03,ala04}.

A key feature is the relation between an avalanche duration $\tau$
and its volume $s(\tau)$. In general, a scaling $s(\tau) \sim
\tau^\gamma$, may be expected, with exponent $\gamma$ determined as
follows. Consider a region of lateral size $\ell$ swept over by an
avalanche. Its vertical extent is related to $\ell$ by the {\em local}
roughness exponent, $w \! \sim \! \ell^{\chi_{\rm loc}}$
\cite{lop97,dub99}, with volume $s \! \sim \! \ell^{d+\chi_{\rm
    loc}}$. The avalanche is driven by a higher capillary pressure
in the moving region as compared to other parts of the front. Since
locally the values of $p_c$ are independent random quantities, the
excess driving force (the difference of typical $p_c$ and the large
fluctuation) and velocity decrease with size as $v \sim
\ell^{-d/2}$. Therefore, the avalanche sweeping time scales as $\tau =
w/v \sim \ell^{\chi_{\rm loc}+d/2}$ and leads to the relation $s \sim
\tau^\gamma$ with
\begin{equation}
\gamma = \frac{\chi_{\rm loc} \! + \! d}{\chi_{\rm loc} \! + \! d/2}.
\label{avalexp}
\end{equation}
For $d \! = \! 1$ we have $\chi_{\rm loc} \! = \! 1$ \cite{dub99} and
expect $\gamma \! = \! 4/3$. In $d \! = \! 2$ we would expect using
$\chi_{\rm loc} \! = \! \chi = \! 3/4$ \cite{3d} $\gamma \! = \!
11/7$.

To see the scaling of velocity fluctuations $\Delta v$ we first relate
the average avalanche velocity ${\bar v}_{\rm ava}$ to the overall
velocity $\bar{v}$ by the relative size of moving parts, ${\bar
  v}_{\rm ava} \! \approx \! (\bar{h}/ \xi)^d \bar{v}$. Next, we
assume a simple relation $\Delta v_{\rm ava} \! \sim \! \bar v_{\rm
  ava}$ between the average avalanche velocity and its fluctuations,
justified by the slow, intermittent motion. Thus
\begin{equation}
\Delta v \approx
\left(\frac{\bar{h}}{L} \right)^\frac{d}{2} \left(
\frac{\xi}{\bar{h}} \right)^d \; \Delta v_{\rm ava}
\end{equation}
by factors resulting from independence on scales larger than $\bar h$
and the relative fraction of moving parts on smaller scales. The
relation between average avalanche velocity and fluctuations yields in
general, both for spontaneous and forced imbibition
\begin{equation}
\Delta v  \sim
\left(\frac{\bar{h}}{L} \right)^{\! \frac{d}{2}}
\left( \frac{\xi}{\bar{h}} \right)^d \; \bar v_{\rm ava}
\sim 
\left( \frac{\bar{h}}{L} \right)^{\! \frac{d}{2}} \; \bar v.
\label{imbifluc} \end{equation}
In spontaneous imbibition with rising height $\bar v \! \sim \!
1/\bar h$ and $\Delta v \! \sim \! \bar v^{1 \! - \! d/2} L^{-d/2}$.
Eq.~(\ref{imbifluc}) thus predicts that the fluctuations depend on the
geometry (via $\bar{h}$ and $L$) and on the average velocity. 

\begin{figure}[!h]
\begin{center}
\includegraphics[width = 4.2cm]{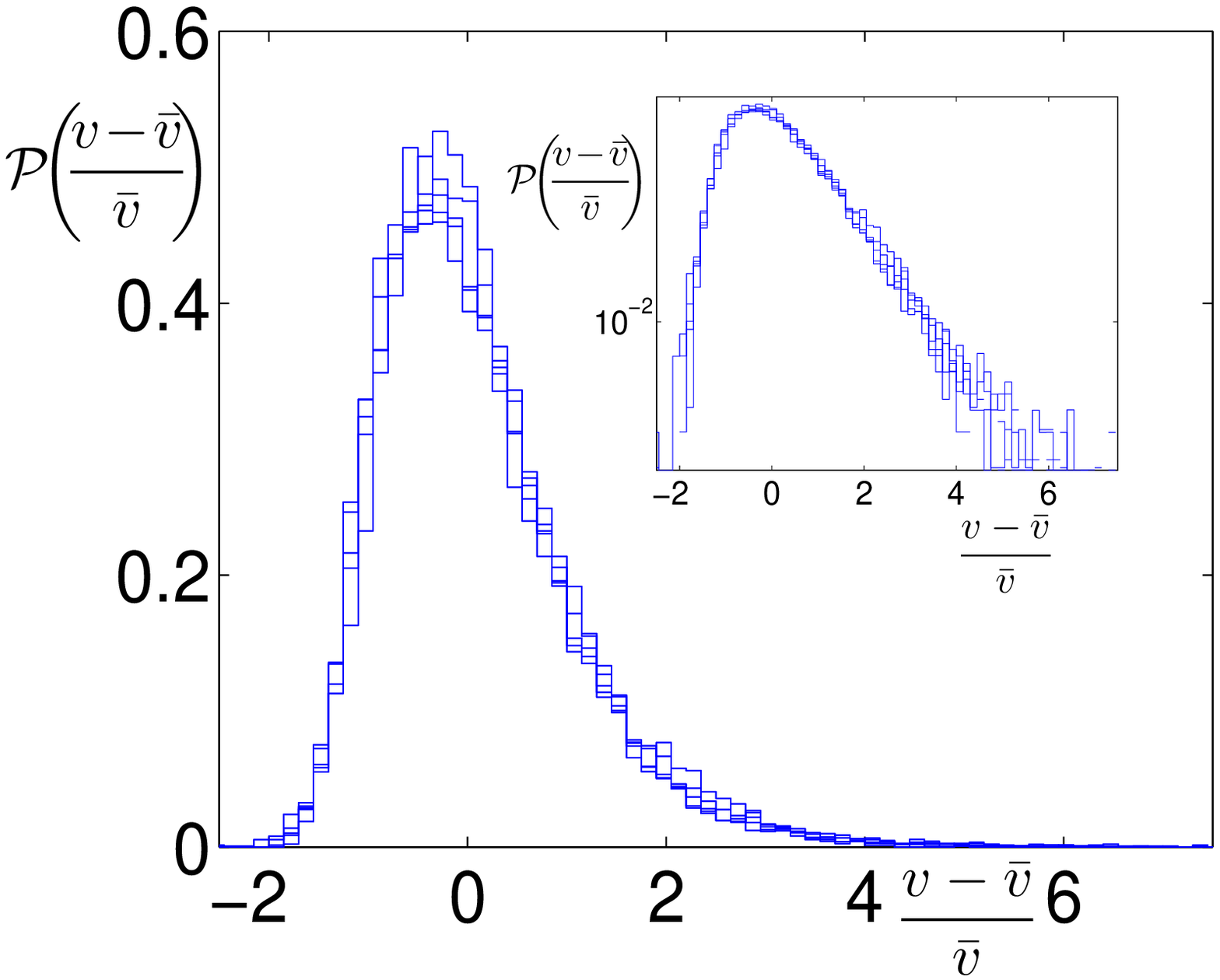}
\includegraphics[width = 4.3cm]{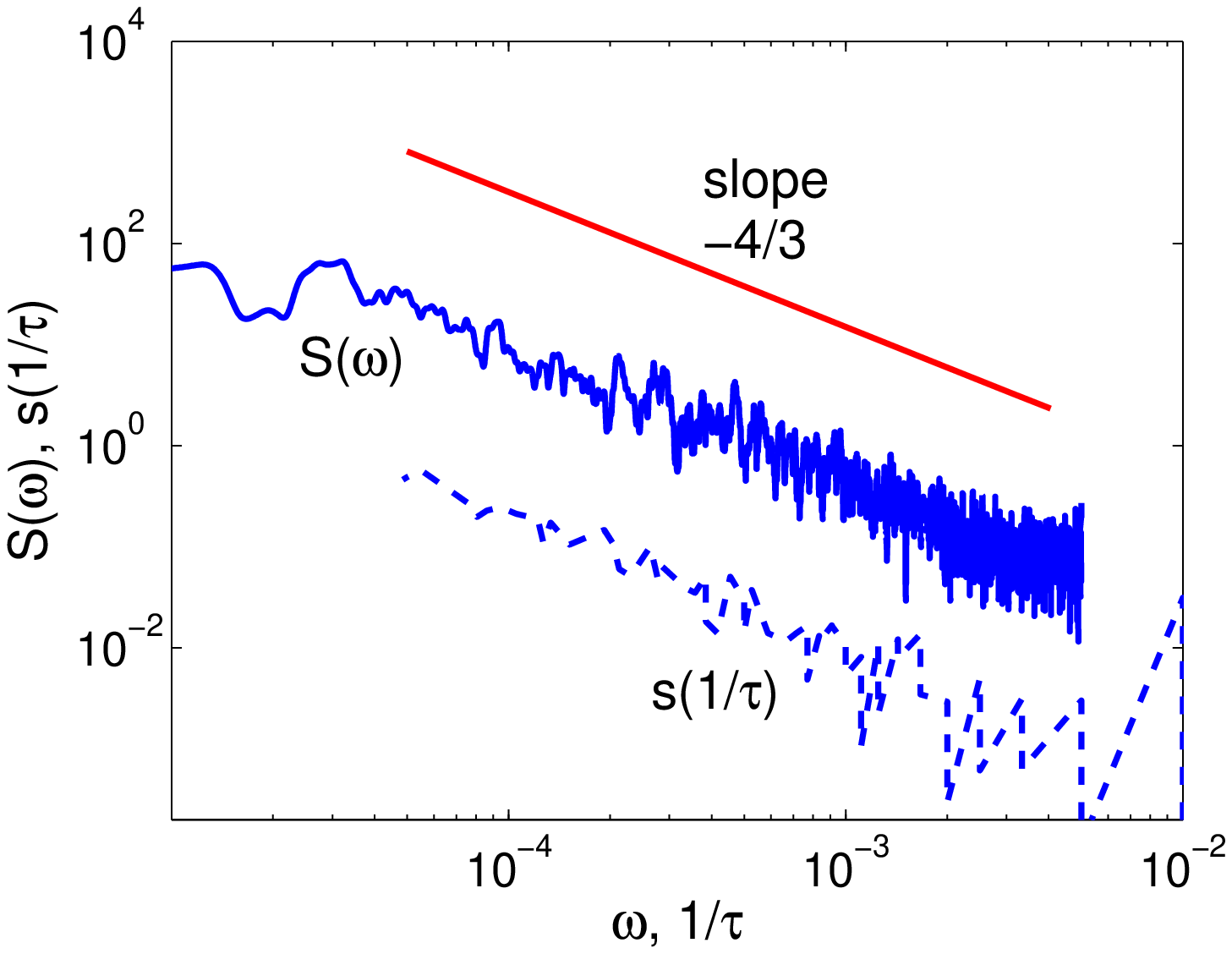}
\end{center}
\vspace{-.5cm}
\caption{{\bf Left}: Rescaled distributions ${\cal P}(\frac{v(t) \! -
  \!{\bar v}}{\bar v})$ for various values of $1/{\bar v} \! \propto
  \! \bar h \! = \! L \! = \! 40, \dots, 1280$. {\bf Right}: Power
  spectrum $S(\omega) \! = \! \langle |\hat v(\omega)|^2 \rangle \!
  \propto \! \omega^{-4/3}$ for $\bar h \! = \! L \! = \! 1280$. Same
  exponent relates average avalanche size $s$ and duration $\tau$.}
\label{fig:pvbar}
\end{figure}

In the left panel of Fig.~\ref{fig:pvbar} we show the distributions of
$P(v(t))$ obtained in phase field simulations for several values of
$\bar{v}$, and {\em constant} square aspect ratio $L = \bar h$: They
can be scaled by an ansatz $P(v) \! = \! \frac{1}{\bar v} {\cal P} \!
\left(\frac{v \! - \! \bar v}{\bar v}\right)$. The shape of $P(v(t))$
resembles a Gumbel distribution of extreme value statistics reflecting
avalanches in regions of maximal capillary pressure, but our data are
too scarce for a careful analysis.

The avalanche motion is reflected in the power spectrum of the
interface velocity, $S(\omega) = \langle |\hat v(\omega)|^2\rangle
\sim \omega^{-\alpha}$. The exponent $\alpha$ may equal the exponent
$\gamma$ discussed previously, if the avalanches have a self-affine
fractal spatiotemporal structure \cite{lau05}. The right panel of
Fig.~\ref{fig:pvbar} compares the power spectrum $S(\omega) \! = \!
\langle | \hat v(\omega)|^2\rangle$ of the mean interface velocity
(solid line) to the average size $s(\tau)$ of avalanches with duration
$\tau$ (dashed line). Both follow power laws, $\omega^{-\alpha}$ and
$\tau^{\gamma}$ with $\alpha \! = \! \gamma \! \approx \! 4/3$, as
predicted by Eq.~(\ref{avalexp}).

\begin{figure}[!h]
\begin{center}
\includegraphics[width = 4cm]{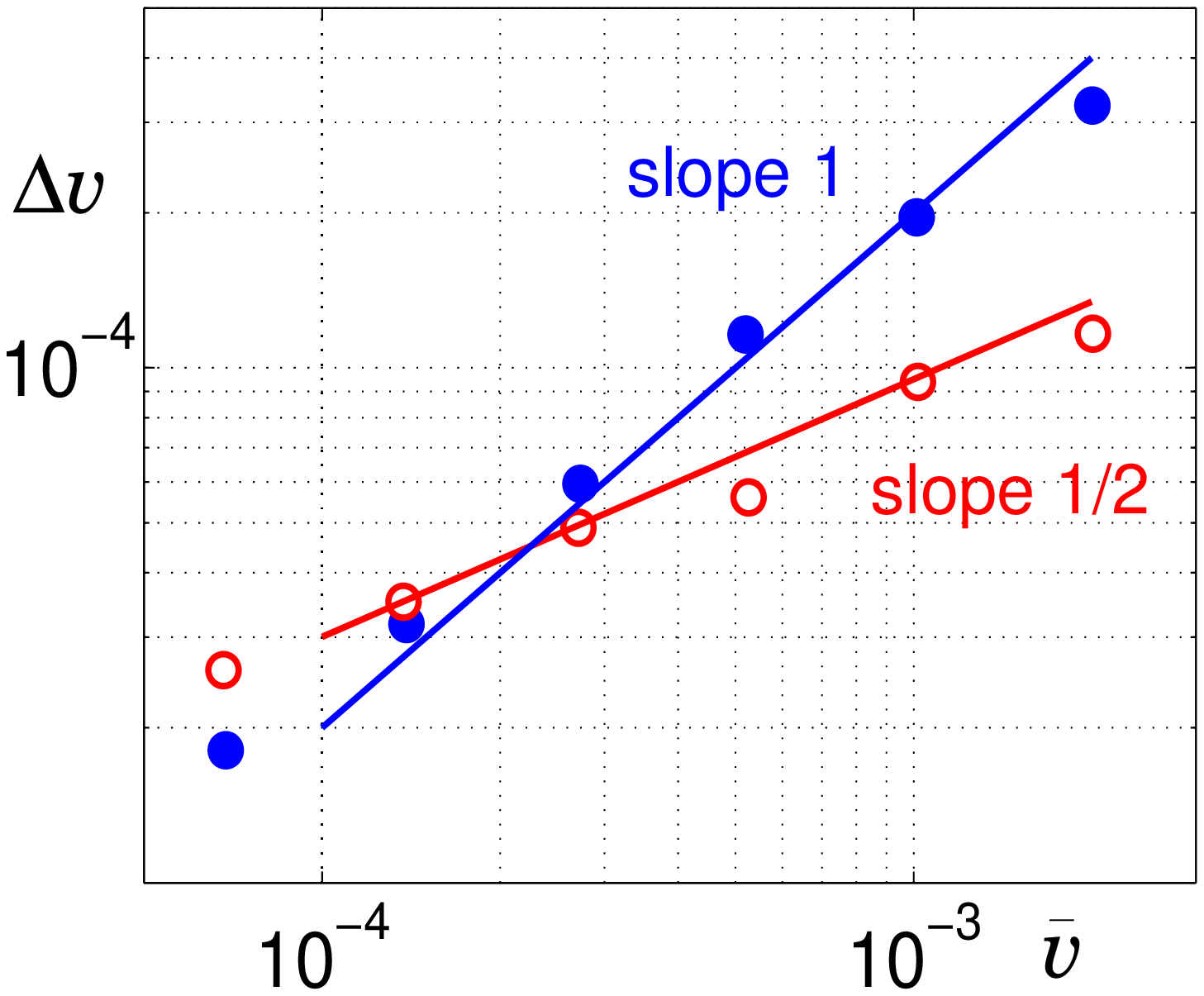}
\includegraphics[width = 4.2cm]{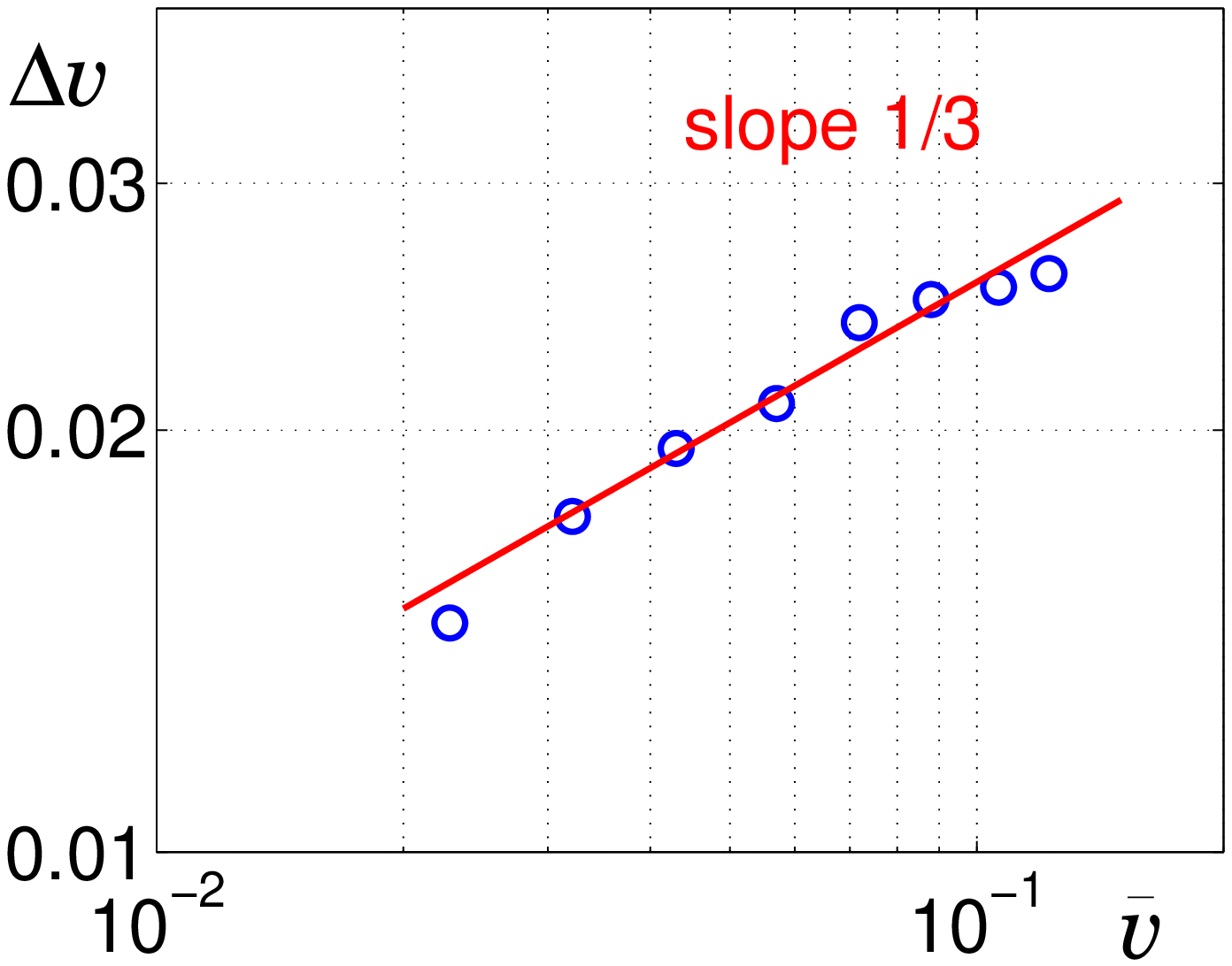}
\end{center}
\vspace{-.5cm}
\caption{{\bf Left}: Fluctuations $\Delta v$ vs.\ mean $\bar v$ for
  constant width $L \! \equiv \! 500$ (open circles) and $L \! = \!
  \bar h$ (full circles), follows $\Delta v \! \sim \! {\bar v}^a$
  with $a \! = \! 1/2$ and $1$ resp., cf.\ Eq.~(\ref{imbifluc}). Lines
  with slope 1/2 and 1 resp.\ guide the eye. {\bf Right}: Same for
  local model of \cite{les93}, with system size $L=10^4$. Line has
  slope 1/3.}
\label{fig:data}
\end{figure}

The left panel of Figure~\ref{fig:data} shows $\Delta v$ vs.\ $\bar v$
for two different choices of the system geometry: $\bar h$ varying,
with $L$ fixed, and with $L \! = \! \bar h$. We observe $\Delta v \!
\propto \! {\bar v}^{a}$ with $a \! = \! 1$ for a square shape and $a
\! = \! 1/2$ for fixed width in agreement with our scaling picture. We
checked also that for $L \! \gg \! \bar h$ the distribution $P(v(t))$
gets sharper with increasing $L$, in agreement with the implication of
the central limit theorem for independent random variables $\Delta v
\! \sim \! L^{-1/2}$ (Eq.~(\ref{imbifluc})).

An interface driven by a force $F$ in a random medium {\em without}
any conservation law obeys different scaling. In the presence of a
critical (zero--temperature, neglecting thermal creep) value $F_c$
separating pinned and unpinned propagation a correlation length $\xi
\! \sim \! \Delta F^{-\nu}$ ensues where $\Delta F \! = \! F \! - \!
F_c$ and $\nu$ is the correlation length exponent. Meanwhile, the
order parameter follows the generic scaling $\bar{v} \! \sim \! \Delta
F^\theta$. In the critical region, when $\Delta F$ is so small that
$\xi \approx L$, the order parameter exponent $\theta$ takes a
nontrivial value smaller than unity. For slightly larger driving
forces, we have $L \! > \! \xi$. In the simulations here we considered
$L/\xi \! = \! O(10)$, so we have $\bar v \! \sim \! \Delta F$ or a
trivial effective $\theta_{\rm eff} \! = \! 1$.

In a system of lateral size $L$ there are $N =(L/\xi)^d$ independent
sub-volumes. Inside these, the fluctuating locally averaged velocities
$v^{\rm loc}_i$ are independent random variables with mean $\bar{v}$
and variance $\delta v^2$. Their average, the instantaneous velocity
$v(t) = \sum_i v^{\rm loc}_i/N$ then has fluctuations measured by
\begin{equation}
\Delta v \sim \delta v/\sqrt{N} \sim \bar
v^{1-d\nu/(2\theta)}/\sqrt{L^d}.
\label{localscaling}
\end{equation}

Numerical simulations are performed with a cellular automaton for the
QEW equation \cite{les93}, $\partial h / \partial t \! = \! \Gamma
\nabla^2 h \! + \! F \! + \! \eta(x,h(x,t))$, where $\eta$ is a
quenched noise term with a bare correlator $\langle \eta({\bf r})
\eta({\bf r}')\rangle \! = \! D \delta({\bf r} \! - \! {\bf r}')$ with
strength $D$. $\Gamma$ is a surface tension similar to $\gamma^*$.
The data are presented in the right panel of Fig.~\ref{fig:data}. They
fit reasonably to the scaling ansatz of Eq.~(\ref{localscaling}) with
$1 \! - \! d\nu/(2\theta_{\rm eff}) \! = \! 1/3$, derived from the
known critical value $\nu \! \approx \! 4/3$ of the one-dimensional
depinning transition \cite{les93} and $\theta_{\rm eff} \! = \! 1$.

To summarize, we have presented a scaling theory for interface
velocity fluctuations. Without detailed measurements of interface
configurations but by the easily accessible fluctuations of the {\em
  average} interface propagation one obtains information about the
universality class of the process. Our arguments show the difference
between dynamics with and without a bulk conservation law behind the
interface. This is also manifest in the power spectrum of $v(t)$,
where the avalanche-like dynamics allows to understand the ensuing
$1/f$-noise. The main argument applies to any dimensionality and is
not dependent on the detailed model. A similar reasoning should extend
to other systems where the global conservation of a quantity is
important. One experimental possibility is vortex penetration dynamics
in (high-$T_c$) superconductors, --- for which there is no known
coarse-grained description \cite{zap01,alt04} --- since the vortex
density is conserved.

{\bf Acknowledgments}. This work has been supported by Deutsche
Forschungsgemeinschaft and the Academy of Finland. We thank Lorentz
Centre (Leiden, NL) and NORDITA (Copenhagen, DK) for kind
hospitality.

\end{document}